\documentstyle[11pt,aaspp4]{article}
\begin{document}
\newcommand{\etal}{et al. }
\newcommand{\tsph}{TreeSPH}
\newcommand{\K}{{\rm K}}
\newcommand{\lya}{Ly$\alpha$ }

\newbox\grsign \setbox\grsign=\hbox{$>$} \newdimen\grdimen \grdimen=\ht\grsign
\newbox\simlessbox \newbox\simgreatbox
\setbox\simgreatbox=\hbox{\raise.5ex\hbox{$>$}\llap
     {\lower.5ex\hbox{$\sim$}}}\ht1=\grdimen\dp1=0pt
\setbox\simlessbox=\hbox{\raise.5ex\hbox{$<$}\llap
     {\lower.5ex\hbox{$\sim$}}}\ht2=\grdimen\dp2=0pt
\newcommand{\simgt}{\mathrel{\copy\simgreatbox}}
\newcommand{\simlt}{\mathrel{\copy\simlessbox}}
\newcommand\junits{{\rm erg\,s}^{-1}\,{\rm cm}^{-2}\,{\rm sr}^{-1}\,
		   {\rm Hz}^{-1}}
\newcommand\iunits{{\rm cm}^{3}\,{\rm s}^{-1}}

\lefthead{Hellsten et. al.}
\righthead{Metal line Observability}

\title{The Observability of Metal Lines Associated with the Lyman-alpha Forest}

\author{Uffe Hellsten and Lars Hernquist\altaffilmark{1}}
\affil{University of California, Lick Observatory, Santa Cruz, CA 95064}
\author{Neal Katz}
\affil{University of Massachusetts, Dept. of Physics and Astronomy, Amherst, MA, 01003}
\and
\author{David H. Weinberg}
\affil{Ohio State University, Department of Astronomy, Columbus, OH 43210} 
\altaffiltext{1}{Presidential Faculty Fellow}

\begin{abstract}
We develop a prescription for characterizing the strengths of metal lines associated with  
\lya forest absorbers (LYFAs) of a given neutral hydrogen column density $N_{\rm HI}$ and 
metallicity [O/H]. This {\em Line Observability Index} (LOX) is line-specific and translates, 
for weak lines, into a measure of the equivalent width. It can be evaluated quickly for 
thousands of transitions within the framework of a given model of the \lya forest, providing
a ranking of the absorption lines in terms of their strengths and enabling model builders to select the 
lines that deserve more detailed consideration, i.e. those that should be detectable in observed 
spectra of a given resolution and signal-to-noise ratio.  
We compute the LOX for a large number of elements and transitions in two cosmological
models of the \lya forest at $z \sim 3$ derived from  hydrodynamic simulations of structure
formation.  
We present results for a cold dark matter universe with a cosmological constant;
an $\Omega=1$ cold dark matter model yields nearly identical results, and we
argue more generally that the LOX predictions are insensitive to the
specific choice of cosmology.
We also discuss how the LOX depends on 
redshift and on model parameters such as the mean baryonic density and radiation field.

We find that the OVI (1032 \AA, 1038 \AA) doublet is the best probe of the metallicity 
in low column density LYFAs $(N_{\rm HI} \approx 10^{14.5} {\rm cm}^{-2})$.
Metallicities down to [O/H] $\sim$ -3 
yield OVI absorption features that should be detectable in current high-quality spectra, 
provided that the expected position of the OVI feature is not contaminated by HI absorption. 
The strongest transitions in lower ionization states of oxygen are OV(630 \AA), OIV(788 \AA), 
and OIII(833 \AA). These absorption lines are all predicted to be stronger than the OVI
feature, but even at redshifts $3-4$ they will have to be observed in the ultraviolet, and they are
extremely difficult to detect with present UV instruments, such as the Space Telescope Imaging
Spectrograph (STIS). At lower redshifts, detection of these lines may be possible in STIS
spectra of the very brightest QSOs, while one may have to wait for next-generation instruments
such as the Cosmic Origins Spectrograph (COS) to detect such lines in a number of
high-redshift QSOs.

The strongest metal lines with restframe wavelength larger than $912 {\rm \AA}$ associated with higher 
column density LYFAs at $z \sim 3$ are CIII (977 \AA) and SiIII 
(1206.5 \AA), which peak at $N_{\rm HI} \sim 10^{17} {\rm cm}^{-2}$. 
Of the lines with rest wavelengths $\lambda_r > 1216 {\rm \AA}$, which can potentially be 
observed redwards of the \lya forest, the CIV(1548,1551) doublet is expected to 
dominate in all LYFAs, regardless of the value of $N_{\rm HI}$.
We argue that CIV and CII absorption may peak in different spatial regions, and that
comparison of single-phase models of the CIV/CII ionization ratios with  observed CIV/CII
column density ratios can lead to an overestimate of the ionization parameter in the
central parts of the absorbers. 

\end{abstract}


\section{Introduction}

Over the past decade, observations have demonstrated that strong \lya forest 
absorbers (LYFAs) generally show associated
metal line absorption (Meyer \& York 1987; Lu 1991; Cowie et al. 1995; Womble et al. 1995;
Songaila \& Cowie 1996, hereafter SC). SC find CIV absorption in nearly all
LYFAs at $z \sim 3$ with neutral column densities 
$N_{\rm HI}\simgt 1.6 \times 10^{15} \, {\rm cm}^{-2}$
and in $\simeq 75 \%$ of systems with $N_{\rm HI}\simgt 3.2 \times 10^{14}$. The observed
CIV column densities are consistent with the absorbers having a mean metallicity 
${\rm [C/H]} \sim -2.5$ and an intrinsic scatter in metallicity of about an order of
magnitude (Rauch et al. 1996; Hellsten et al. 1997). An important question is whether 
or not a chemically pristine or extremely metal poor
population of LYFAs exists, and if so, what the characteristic HI column densities
of this population are. For $\log{N_{\rm HI}} \simlt 14.5$ the associated CIV lines are close to
the detection limits of SC, and it cannot yet be determined if these absorbers 
have the same metallicity distribution as those with higher $N_{\rm HI}$.  
Some theoretical models of early metal enrichment predict that the mean metallicity
declines steadily with decreasing $N_{\rm HI}$ (Gnedin \& Ostriker 1997).

Considerable progress has been made in the theoretical understanding of LYFAs. From the work of
several groups (e.g. Cen et al. 1994; Zhang et al. 1995; Petitjean et al. 1995;
Hernquist et al. 1996; Bi \& Davidsen 1997) 
a picture has emerged in which the LYFAs are interpreted in terms of
``Gunn-Peterson'' absorption in an inhomogenous IGM pervaded by a background ionizing 
radiation field, presumed to originate from QSOs and perhaps young, star-forming galaxies. 
This picture links the \lya forest directly to cosmological structure formation. The 
analysis of cosmological models of the LYFAs consists of producing artificial
spectra by evaluating the absorption properties of the baryonic IGM along lines of sight
through N-body and hydrodynamical simulations of structure formation. These spectra can
then be analyzed in much the same way as observed QSO spectra, and the resulting distribution
functions in $N_{\rm HI}$ and linewidths are found to agree quantitatively with those
observed, lending considerable support to the cosmological LYFA picture
(e.g. Miralda-Escud\'{e} et al. 1996; Dav\'e et al. 1997; Zhang et al. 1997).

The interpretation of the metal line data within the framework of these models
provides a strong test of this picture. If a metal enrichment pattern 
of the baryonic IGM is specified, the properties of selected absorption lines are readily
calculated from the knowledge of densities, temperatures, and UV radiation field along
the lines of sight (Haehnelt et al. 1996; Rauch et al.  1996; Hellsten et al. 1997).  
Until now, the metal lines considered  have been limited to OVI(1032,1038), NV(1239,1243),
CIV(1548,1551), SiIV(1394,1403), and CII(1335). These lines have been chosen because they
have already been identified with LYFAs in real QSO spectra. With the exception
of the OVI doublet they all have rest wavelengths $\lambda _0 > 1216$ \AA{}, which means that 
for LYFAs at sufficiently high redshifts, they appear redward of the \lya forest and hence
are relatively easy to detect. Instead of making an a priori selection of a few metal 
lines to include in a cosmological model for LYFAs, a more satisfactory approach would be to 
make the models {\em predict} which metal lines, out of thousands of candidates, should be
observable (in spectra of a given resolution and S/N) in LYFAs of different 
column densities. Such an approach would allow a comprehensive and swift selection
of lines that deserve a more detailed treatment, depending on the specific purpose of the 
modeling.  It would provide a sharper test 
of the model and place more stringent constraints on the metallicities and UV radiation 
spectrum, by making it possible to compare the complete list of metal lines predicted to be 
observable to the list of lines actually detected in QSO spectra, as well as the observed 
relative strengths of these lines. 

This paper presents and applies a technique for implementing this comprehensive, a priori
approach to metal line absorption predictions in cosmological models.
In Section 2 we define a line observability 
index, which can easily be evaluated for a database of metal lines, within a particular model
for LYFAs. This index is a measure of column density and hence allows the lines to be sorted
by strength. It also predicts which lines should be detectable in LYFAs for a wide range of column 
densities and metallicities. In Section 3 we apply this technique to two specific
cosmological models of the \lya forest, based on hydrodynamic simulations of
structure formation.
In Section 4 we discuss the dependence of the LOX on model parameters such as the mean
baryon density and radiation field and estimate how it changes with redshift, and we 
argue that the LOX has only a very weak dependence on the underlying cosmology.
We discuss the results and summarize the conclusions in Section 5.

\section{The Line Observability Index}  
Detection limits for absorption lines can be conveniently expressed in terms of 
the equivalent width  

\begin{equation}
W_{\lambda} \, \equiv \, \int _0 ^{\infty} \left( 1-\frac{F_{\lambda}}{F_{c\lambda}} \right) 
d \lambda,
\end{equation}

\noindent where $F_{\lambda}/F_{c\lambda}$ denotes the received energy flux per wavelength 
interval, normalized to the continuum level. 
For absorption lines that are weak enough to be optically thin,
the rest frame equivalent width $W_{\lambda r}$ is proportional to the column density $N$
of absorbing atoms along the line of sight
 
\begin{equation}
W_{\lambda r} \, = \, \frac{\pi \, e^2}{m_e c^2} \, N \, {\lambda _r}^2 \, f,
\label{lincog}
\end{equation} 

\noindent where $f$ and $\lambda _r$ are the oscillator strength and rest frame wavelength 
of the line considered. The other symbols have their usual meaning. Lines that
satisfy equation (\ref{lincog}) are said to be on the linear part of their
curve-of-growth. (For a derivation of this relation see, e.g., Spitzer 1978.)
We will label an absorption line produced by an element $Z$ in ionization
stage $i$ with rest transition wavelength $\lambda$ and oscillator strength $f$
as $Z_{\lambda , f}^i$.  

For a metal absorption line of metallicity 
$[{Z/H}] \, \equiv \, \log{(n_Z/n_{\rm H})} \, - \, \log{{(n_Z/n_{\rm H})}_{\odot}}$ associated with 
a LYFA of hydrogen column density $N_{\rm HI}$ we now define the following quantity:
\begin{equation}
{\rm LOX}(Z_{\lambda , f}^i,[Z/H],N_{\rm HI}) \, \equiv \, -17.05 \, + \, 
\log{N_{\rm HI}} \, + \, \left[\frac{Z}{\rm H} \right] \, + \log{(f\, {\lambda}^2)} \,
+ \log{{\left(\frac{n_Z}{n_{\rm H}}\right)}_{\odot}} \, + \, 
\log{\left(\frac{x_Z^i}{x_H^0}\right)}, 
\label{lox}
\end{equation} 
which we will call the {\em line observability index}.
The inspiration for this definition comes from rewriting equation (\ref{lincog}) 
with the column density of the absorbing metal ions expressed in terms of $N_{\rm HI}$,
[Z/H], and the solar ratio of the number densities of metal and H atoms ${(n_Z/n_{\rm H})}_{\odot}$.
The model-dependent ionization corrections are contained in the last term, which is 
a mean value of the ratio of the ionization fractions $x_Z^i = n_Z^i/n_Z$ of $Z^i$ and $H^0$ 
in the region where the \lya absorption arises. 
Equation (\ref{lox}) assumes units of ${\rm cm}^{-2}$ for $N_{\rm HI}$ and \AA{} for $\lambda$;
the choice of additive constant then implies 
${\rm LOX} \, = \, \log{(W_{r \lambda}/\, 1\, m{\rm \AA})}$ for weak lines.
The LOX can thus be used to rank metal lines in terms of strength, and it can be directly 
compared to the detection limit in spectra of a given quality.  
Computation of the LOX as a function of $[Z/H]$ and $N_{\rm HI}$ requires
a model of the densities and temperatures within \lya forest absorbers, which
together with the ionizing background radiation field $J_\nu$
determine the ionization term $x_Z^i/x_H^0$ as a function of $N_{\rm HI}$.

\section{LOX for 2249 candidate metal lines}
We have applied the LOX technique to two different cosmological simulations. One is
the standard CDM simulation described in Dav\'{e} et al. (1996). A box of length 22.222 comoving Mpc 
containing baryonic and dark matter is evolved from $z=49$ to $z=2$ using TreeSPH 
(Hernquist \& Katz 1989), assuming a CDM spectrum of density fluctuations with the parameters
$\sigma_{8h^{-1}} = 0.7$, $\Omega = 1.0$, $h = 0.5$, $\Omega_b h^2 = 0.0125$,
and $n = 1$, where $h \equiv {\rm H}_0/100\;$km/s/Mpc and $n$ is the slope of 
the initial perturbation spectrum.  The other simulation, a flat, nonzero-$\Lambda$
model, also assumes a CDM spectrum of density fluctuations,
with $\sigma_{8h^{-1}} = 0.8$, $\Omega = 0.4$, $\Lambda = 0.6$, $h = 0.65$,
$\Omega_b h^2 = 0.0125$, and $n = 0.93$.  
The standard model is inconsistent with the amplitude of CMB anisotropy measured by
COBE (Smoot et al. 1992), but the normalization of the $\Lambda$ model
(computed using the CMBFAST program of Seljak \& Zaldarriaga 1996)
is matched to the 4-year COBE data.
For further details regarding the simulation method, see Katz, Weinberg, \& Hernquist (1996a).

The analysis of the two cosmological models yielded very similar results. The LOX is a local property, 
depending on the conditions within individual density peaks along the line of sight, which are
insensitive to the underlying cosmology. Hence, in the following we will only describe the 
calculations for one of the models, the $\Lambda$CDM model, and we will focus on results
at $z=3$.
A spatially uniform photoionizing background radiation field is imposed, with the shape 
computed by Haardt \& Madau (1997, in preparation). They calculate the modification due to 
absorption and re-emission by the IGM of an input radiation field with $J_{\nu} \propto \nu ^{-1.8}$
for $\lambda < 1050 {\rm \AA}$, $J_{\nu} \propto \nu ^{-0.9}$ for $1050 {\rm \AA} \leq \lambda
\leq 2200 {\rm \AA}$, and $J_{\nu} \propto \nu ^{-0.3}$ for $\lambda > 2200 {\rm \AA}$.
This radiation field, at $z=3$, has a larger drop above the ${\rm He}^+$ absorption edge at 
4 Rd than the previously published Haardt \& Madau spectrum (1996), by roughly a factor of three,
and it is therefore in better agreement with the relatively large number of SiIV systems
observed at that redshift (see the discussion in Hellsten et al. 1997). 

To evaluate the ionization correction term in expression (\ref{lox}) we need to know the
typical volume densities and temperatures in the absorbers for different values of
$N_{\rm HI}$. To find the relation between these quantities, we generated artificial
\lya absorption spectra along 480 lines of sight through the simulation box at $z=3$,
using the method described in Hellsten et al. (1997).
The lines of sight were selected to include a relatively large number of the high-column 
density absorbers in order to get good statistics for these rather uncommon systems. 
We use artificial spectra along random lines of sight to fix the overall
amplitude of the background radiation field, scaling $J_\nu$ by a 
constant factor so that the mean \lya flux decrement in the
simulated spectra matches the value $D_A=0.36$ found by 
Press, Rybicki, \& Schneider (1993; see further discussion by
Rauch et al.\ 1997).

Figure \ref{figcolnh} shows results from an analysis of 1456 HI absorption features detected 
in these artificial spectra. Lines are identified by a threshold criterion in 
the following way. First, we define the velocity interval from a position where the HI optical depth
$\tau _{\alpha}$ increases above a critical value $\tau _c = 1.2$ until it drops below 
that value again as belonging to an HI absorption feature. Next, we augment this interval on both
sides until either $\tau _{\alpha}<0.1$  or a local maximum
in $\tau _{\alpha}$ is encountered. This procedure separates weakly blended ``W-shaped''
absorption features, if $\tau _{\alpha}<\tau _c$ at the central local minimum.

The HI column density of the absorber is found by integration of $\tau _{\alpha}$ 
over the velocity interval, i.e.

\begin{equation}
N_{\rm HI} \, = \, \frac{m_e c}{\pi e^2 f_{\alpha} \lambda _{\alpha}} \, \int _{\Delta v}
\tau _{\alpha}(v) \, dv \, = \, 7.44 \times 10^{11} \, {\rm cm}^{-2} \, \int_{\Delta v}
\tau _{\alpha}(v) \, dv,
\end{equation}
where $v$ is measured in km/s in the rightmost expression. 
The crucial question for our purposes is the relation between $N_{\rm HI}$ and 
the typical volume density of gas in
the absorbing region. Because of peculiar motions and Doppler broadening,
the absorbing HI atoms in a given velocity bin originate at slightly different spatial 
locations. The code we developed to generate the artificial spectra also determines the
{\em mean} total hydrogen volume density $n_{\rm H}(v)$ in the regions from which the HI atoms 
originate. In Figure \ref{figcolnh} we plot $n_{\rm H}$ vs. $N_{\rm HI}$, where $n_{\rm H}$ 
is the maximum value of $n_{\rm H}(v)$ in the velocity interval $\Delta v$
of that absorption feature. In the spectra, the location of this density peak is close to the
location of the maximum of $\tau _{\alpha}(v)$, as one would expect.

As can be seen in Figure \ref{figcolnh}, there is a well-defined correlation between
the HI column density of a LYFA and the typical total volume hydrogen density of the absorbing
region. The thick solid line is the power-law fit 

\begin{equation}
\log{n_{\rm H}}=-14.8 + \log{\frac{\Omega_b h^2}{0.0125}} + 0.7 \log{N_{\rm HI}},
\label{nnhfit}
\end{equation}

\noindent which we will use to evaluate the LOX. 

Relation (\ref{nnhfit}) depends on the value of $\Omega_b h^2$, i.e., the total mean baryonic
density.  Increasing the value of $\Omega_b h^2$ will shift the curve upward if
$N_{{\rm HI}}$ is held constant (to satisfy the mean flux decrement constraint) by adjusting the
intensity of the UV radiation field. The indicated scaling with $\Omega_b h^2$ in 
(\ref{nnhfit}) is only approximate, because it neglects effects of the temperature 
dependence on $\Omega_b h^2$ (Croft et. al. 1997). We will assume $\Omega_b h^2=0.0125$ 
throughout this paper, except for the discussion in Section 4. 
While a similar correlation between $n_{\rm H}$ and $N_{\rm HI}$ holds at other redshifts,
the constant of proportionality is different because of the evolution in the radiation
field and the IGM structure, in particular the dilution of the cosmic mean density
by the expansion of the universe. The $n_{\rm H}$-$N_{\rm HI}$ correlation for the
SCDM model is very similar, with a slope of about 0.73 instead of 0.7, and the resulting
line observability indices are nearly identical for the two relations.

Also shown in Figure \ref{figcolnh} is the mean total hydrogen density at $z=3$ in
this model. It is seen that the model predicts that LYFAs with 
$\log{(N_{\rm HI})} \simlt 14$ at $z=3$ arise from underdense regions
(similar to the result of Zhang et al. 1997).

The relation between density and temperature is slightly more complex. Figure~\ref{figtnh}
shows a plot of $n_{\rm H}$ versus mean temperature for the same density peaks as
those shown in Figure \ref{figcolnh}. For low densities a large fraction of the
peaks follow the relation $T \propto n_{\rm H}^{0.65}$. 
At densities $n_{\rm H} \simlt 10^{-4} {\rm cm}^{-3}$, the density-temperature
correlation arises mainly from the interplay between photoionization heating
and adiabiatic cooling due to the expansion of the universe.
Low density regions have lower neutral fractions, hence lower photoionization
heating rates, and they expand more rapidly, thus experiencing stronger
adiabatic cooling.  Some of the peaks lie above this temperature-density relation,
at $T \sim 10^{4.5-5} {\rm K}$, because of moderate shock heating.
Peaks with $\log n_{\rm H} \simgt -3.6$ have large enough densities for radiative
cooling to become significant, and the temperature is roughly determined by the
equilibrium between radiative cooling and photoionization heating. (For a 
more detailed discussion of the density-temperature relation in cosmological simulations, see,
e.g., Croft et al. 1997; Hui \& Gnedin 1997; Zhang et al. 1997). 
For our LOX calculations we will adopt the relation
defined by the solid line in Figure \ref{figtnh}. This relation is again very similar
to that in the SCDM model, although the latter shows a slightly larger scatter at
lower densities. Because photoionization is the main source of ionization at such
low densities, the LOX is insensitive to the exact value of the temperature.

For an arbitrary metal line $Z_{\lambda,f}^i$ associated with a LYFA with neutral
column density $N_{\rm HI}$ and metallicity [Z/H], we can now evaluate the 
ionization correction term in (\ref{lox}) using the photoionization code CLOUDY 90 
(Ferland 1996).  The inputs to CLOUDY are the hydrogen volume density and the 
temperatures from the relations discussed above, as well as the UV radiation field. 
For systems with $\log{N_{\rm HI}} \simgt 16$, continuum HeII absorption affects the
radiation field in the central regions of the absorbers. 
We include this effect in an approximate way, by solving self-consistently for the radiative 
transfer through a plane parallel slab with HI column density $2 \times N_{\rm HI}$, 
illuminated from both sides. We then use the volume averaged values of the ionization
fractions, which should roughly be equal to those in the central region of an absorber
with HI column density $N_{\rm HI}$. Including this HeII-shielding effect only changes 
the predicted LOX values slightly.

The results in Table 1 are produced by evaluating the LOX for the 2249 transitions
in the compilation by Verner, Barthel, and Tytler (1994) and then sorting the lines according to 
the observability index. The strongest lines are shown for five different values
of the HI column density, 
$\log{N_{\rm HI}}=13$, 14, 15, 16, and 17. We show, for  $\log{N_{\rm HI}} >14$, only the
ten strongest lines with $\lambda _r < 912 {\rm \AA}$. These extreme ultraviolet
lines are numerous, but in spectra of the faint QSOs at $z \sim 3$ they are extremely 
difficult to detect, even with state of the art instruments such as the Space Telecope 
Imaging Spectrograph, installed on the Hubble Space Telescope (HST) in February 1997 
(e.g. Danks et al. 1996).  
At somewhat lower redshifts, the prospects should be better for observing these lines 
in STIS spectra of a few very bright QSOs, such as HE 2347+4342 
($z$=2.88, Reimers et al. 1997), but the observations have yet to be made.
With the Cosmic Origins Spectrograph (COS), scheduled to be installed on the HST in 2002
(Green et al. 1997), detection of these important lines may finally become possible for
a larger number of QSOs at high redshift.
 
A metallicity of ${\rm [O/H]}=-2.5$ has been assumed in Table 1, and we use the solar 
metallicity values from Grevesse \& Anders (1989). We evaluate the
LOX for solar {\em relative} abundances, i.e. [Z/O]=0. It can then be scaled to any other
relative abundance pattern a posteriori. For example, if [O/C]=0.5, as observed in some
low metallicity systems (e.g. Wheeler et al. 1989), we have to add 0.5 to the LOX values
for O lines, (i.e. use [O/H]=-2) in order to compare them to the C lines for [C/H]=-2.5.

Before discussing the results in Table 1, let us estimate detection limits
in present day, high signal-to-noise ratio (S/N) HIRES spectra. For a line with rest frame wavelength 
$\lambda_r$, the rest equivalent width threshold for an
$N_{\sigma}$-sigma detection is

\begin{equation}
{\rm W}_r(N_{\sigma}) \, = \, 
\frac{N_{\sigma} \lambda_r \sqrt{N_{\rm pix}}}{\rm S/N}(\Delta v_{\rm pix}/c)\,
= \, \frac{N_{\sigma} \sqrt{N_{\rm pix}}\, \Delta \lambda _{\rm pix}}{{\rm S/N}\, (1+z_{\rm abs})},
\label{declimit}
\end{equation}

\noindent where $N_{\rm pix}=7$ is the number of pixels used for the equivalent width 
detection limit and $\Delta v_{\rm pix}$ ($\Delta \lambda _{\rm pix}$) is the velocity 
(wavelength) spread per pixel (Churchill 1997). 
For the spectrum of Q1422+231 obtained by SC, 
$\Delta \lambda = 0.06 {\rm \AA}$ and $z_{\rm abs} \approx 3$. Hence lines with
${\rm LOX} \simgt (2.3-\log{\rm S/N})$ should be detectable at the 5-sigma level in that spectrum.  
(Recall from \S 2 that $W_r = 1{\rm m\AA} \times 10^{\rm LOX}$ for weak lines.)
The S/N per pixel is roughly 110 outside the \lya forest and 45 within the forest, 
which translates into a minimum detectable LOX of approximately 0.26 outside the forest and 
0.65 within the forest. The integration time for this spectrum was 8.2 hrs. If the
QSO is observed for 3-4 nights the S/N ratio could be roughly doubled, so a practical
absolute lower limit for observable lines would be LOX $\simgt 0$. 

Looking at Table 1, the OVI(1032,1038) doublet is predicted to be a dominant
metal line feature associated with LYFAs with $\log{N_{\rm HI}}=13$, 14, and 15. For
[O/H]=-2,  OVI(1032) becomes observable for $\log{N_{\rm HI}} \simgt 13$ and tops the
list for more than two orders of magnitude in $N_{\rm HI}$. Because this OVI doublet has
rest wavelengths less than 1216 \AA, the absorption features will always be located in
the \lya forest where they are subject to heavy blending and blanketing by HI 
lines, making them difficult to identify, especially in high redshift spectra. 

A family of strong absorption lines from lower ionization states of oxygen is seen to be present
for the higher column density LYFAs. These lines all have $\lambda _r \leq 833 {\rm \AA}$, and
cannot (for $z \simlt 4-5$) be observed in the optical. They present a challenge for the next
generation of ultraviolet spectrographs.

For lines
with $\lambda_r > 1216 {\rm \AA}$, the CIV(1548,1551) doublet is seen to be dominant over the 
entire range of HI column densities, 
a remarkable result, given that this range in $N_{\rm HI}$ spans three orders of magnitude 
in volume density and hence photoionization parameter. 
Only for HI column densities 
close to that of Lyman limit systems ($\log{N_{\rm HI}} \simgt 17.2$) do other metal lines, 
such as the SiIV(1394,1403) doublet, attain strengths comparable to that of CIV. This 
prediction is consistent with the fact that the first metal lines found to be associated with 
LYFAs were CIV lines, and that CIV systems are the best studied and most numerously observed 
metal line systems to date. 

The lines with $\lambda _r > 912 {\rm \AA}$ reaching the largest LOX values are CIII(977) 
and SiIII(1206). In the high column density LYFAs these lines are predicted to be so strong 
that the prospects of detecting them in the \lya forest should
be reasonable, and they have indeed been observed in a few LYFAs in the Q1422 spectrum
(SC 96). These lines deserve a more thorough investigation with artificial
absorption spectra. A good way of testing simulation predictions for 
high column density LYFAs would be to compare CIII absorption in real spectra to the CIII features 
in a model with [C/H] determined observationally from the corresponding CIV features.

Figures 3 and 4 show graphical representations of the $N_{\rm HI}$-LOX relations for 12 lines 
selected from Table 1. In low density regions, the LOX
is insensitive to the scatter in gas temperature, because photoionization is much more important than
collisional ionization and because the recombination coefficients are not very temperature sensitive.
A more complicated issue is what would happen if the {\em mean} temperature of the low density 
gas were different in the model, for example because of a different assumption about the
heat input during the reionization period (Hui \& Gnedin 1997; Rauch et al. 1997). 
The effect on the LOX is small in this case also, as we will argue in detail in Section 4.

Figure 3 contains lines from three different ionization states of carbon and silicon.
The CIV(1548), CIII(997), and CII(1335) lines are expected
to be observable in absorbers with $\log{N_{\rm HI}} \simgt 14$, 14.9, and 15.8, respectively.
These values are obtained assuming a resolution and S/N similar to that of SC's spectrum
of Q1422+231, and a metallicity of [O/H]=-2.5. It is straightforward to predict these
values for spectra of other resolutions and signal-to-noise ratios by means of equation 
(\ref{declimit}), and for other metallicities by shifting the LOX-curves upwards by an amount 
$\Delta$(LOX) = [O/H]+2.5 (or alternatively sliding the horizontal detection limit lines
downwards by the same amount). For example, for CIV to be detectable in a LYFA with
$N_{\rm HI}=10^{13} \, {\rm cm}^{-2}$ in the SC spectrum, one can estimate from Figure 3
that [O/H]  would have to be  $\simgt -0.8$. 

The diagonal arrows on Figure 3 indicate how the curves will shift if $J_{\nu}$ is 
multiplied by a factor of 2 (solid arrow) or if $\Omega _b h^2$ is changed from 0.0125 to
0.02 and $J_{\nu}$ is adjusted to maintain the observed $D_A$ (dashed arrow). 
The direction and magnitude of these arrows can be derived from approximate scaling laws, 
as will be discussed in Section 4.

The modeling of the ionization correction term in the LOX is based on a single-density model, 
assuming that most of the absorption takes place at or near the maximum of the local density 
peak associated with a LYFA. This is certainly true for HI. As long as the volume density of 
a species $Z^i$ increases with total density (i.e. the LOX increases with $N_{\rm HI}$) this
assumption holds. In situations where $Z^i$ decreases with total density this 
approximation becomes worse. For example, in our artificial spectra we sometimes see that the
CII(1335) feature associated with a high column density absorber ($\log{N_{\rm HI}} \sim 16.5$)
is centered on the same velocity as HI, whereas the corresponding CIV (or OVI) absorption is seen
as shoulders that are offset relative to the HI feature. This occurs because the CIV density profile  
does not follow that of the total density. Therefore, our maximum density assumption causes us to
underestimate the LOX values on the decreasing parts of the curves in Figures 3-4. They are lower 
limits to the true LOX values, and it is necessary
to do a full generation and analysis of artificial spectra if one wants to use, say, 
observed CII/CIV ratios to infer ionization properties of the IGM. Direct comparison to
the LOX values in Figure 3 would overestimate the photoionization parameter in the central
region of the LYFA.

Figure 3 also shows the silicon lines SiIV(1394), SiIII(1206), and SiII(1260). Here, 
the SiIV and SiIII features are observable only for $\log{N_{\rm HI}} \simgt 15.8$, and
the SiII feature for $\log{N_{\rm HI}} \simgt 16.3$. In addition, the SiII line falls
redward of the \lya forest only for LYFAs close to the \lya emission peak.
Clearly, for most LYFAs these silicon lines are not as promising for probing
ionization conditions as the carbon lines discussed above. 

Lines from the family of very strong O transitions OIII(833), OIV(788), OV(630), and 
OVI(1032), as well as NV(1239) and SVI(1063), are displayed in Figure 4.  OVI is by far the 
strongest 'high ionization' line with $\lambda _r > 912 {\rm \AA}$,
and it is observable in the entire interval $13.6 \simlt \log{N_{\rm HI}} \simlt 15.8$.
For systems with $\log{N_{\rm HI}} \approx 14.5$, OVI should be detectable in the
SC spectrum, provided ${\rm [O/H] \simgt -3}$. 
In our model, the OVI is produced mainly from photoionization, and the LOX is 
not very sensitive to the assumed temperature. 
With present day spectra, OVI is the
best and only candidate for probing the metallicity of the IGM in regions  
close to the mean density. As we already mentioned, observations of the OVI
lines are hampered by the high line density within the \lya forest. 
Low column density lines in cosmological simulations are usually broadened by
Hubble flow and peculiar velocities rather than thermal motions, so the OVI
lines may not necessarily be distinguishable by low $b$-parameters.
In individual cases, however, it should at least be possible to place upper limits to the metallicity
of the systems that happen not to have strong HI contamination at the expected positions
of the OVI features. 

The NV(1239,1243) is predicted to be weaker than the OVI doublet and probe a somewhat
denser part of the IGM. If [N/C] is as low as -0.7 as suggested by Pettini et al. (1995)
the NV doublet will be observable only in LYFAs with relatively high metallicity, roughly
$[{\rm O/H}] \simgt -1.8$. This seems consistent with the fact that SC observed NV 
in only one out of six candidate absorbers in their Ly$\beta$ selected sample.

Towards lower redshifts, where the \lya forest begins to thin out,
it should become increasingly easy to identify OVI absorption lines by their coincidence
with corresponding \lya and Ly$\beta$ absorption, at least in a 
statistical way.  Though somewhat arduous, this type of investigation with
OVI is probably the most promising route to testing the trend between
metallicity and HI column density predicted by some models of early 
IGM enrichment (e.g., Gnedin \& Ostriker 1997).

\section{LOX dependence on model parameters}

\newcommand\Nh{{N_{\rm HI}}}
\newcommand\Nhp{{N^\prime_{\rm HI}}}
\newcommand\Nhpp{{N^{\prime\prime}_{\rm HI}}}

Figures 3 and 4 present LOX results for a single redshift and a single set of
model parameters: $\Lambda$CDM at $z=3$, with $\Omega_b h^2 =0.0125$ and
the Haardt-Madau (1997) UV background spectrum scaled in intensity to
give an HI photoionization rate $\Gamma_{\rm HI}$ such that $D_A=0.36$ in the simulated spectra.
However, we can estimate the effects of changing some of these parameters
by simple scaling arguments because the dynamics of the gas in the density
regime relevant to LYFAs is driven by the gravity of the dark matter,
making the spatial distribution of the gas insensitive to changes in
$\Omega_b$, $J_\nu$, or the IGM temperature.
We can therefore compute the effects of such changes on the
$\Nh$ vs. $n_{\rm H}$ correlation (Figure~\ref{figcolnh}), and we can
compute the effect on LOX values through equation~(\ref{lox}) if we 
make the approximation that the ionization fractions $x_Z^i$
depend only on the ionization parameter $\Gamma /n_{\rm H}$.  Because 
recombination rates vary only slowly with temperature and the 
range of temperatures in LYFAs is much smaller than the range
of densities, this is a good, but not a perfect, approximation.

Consider first the effect of changing $J_\nu$ by a constant factor $q$.
This change in isolation would alter the mean flux decrement $D_A$,
which we have thus far adopted as an observational constraint on 
the model, but current observational uncertainties in $D_A$ allow
some range in $J_\nu$ (compare, e.g., Zuo \& Lu 1993 to
Press et al.\ 1993 and Rauch et al.\ 1997), and comparison of 
simulated and observed HI column density distributions favors
a somewhat higher value of $J_\nu$ than we have adopted here
(Dav\'e et al.\ 1997; Gnedin 1997a).
We assume that the shape of $J_\nu$ does not change, so that
the photoionization rates $\Gamma$ for all species change by
the same factor.  An absorber that previously had column density
$\Nhp$ now has a column density $\Nh=q^{-1}\Nhp$ ---
raising $J_\nu$ lowers the hydrogen neutral fraction and
therefore lowers $\Nh$.  The density $n_{\rm H}$ in this absorber
is the same as before, so the ionization parameters $\Gamma/n_{\rm H}$
are the same as those in an absorber that previously had column
density $\Nhpp = q^{-1.43}\Nhp = q^{-0.43}\Nh$.
We have used the correlation $n_{\rm H} \propto \Nh^{0.7}$ from 
Figure~\ref{figcolnh} to convert a factor $q^{-1}$ change in
$n_{\rm H}$ to a factor $q^{-1.43}$ change in $\Nh$.  Assuming that
$\log (x_Z^i/x_H^0)$ depends only on $\Gamma/n_{\rm H}$, we see
from equation~(\ref{lox}) that
\begin{equation}
{\rm LOX}(qJ_\nu,\log \Nh) = {\rm LOX}(J_\nu,\log \Nh-0.43\log q) +0.43\log q,
\label{scal1}
\end{equation}
where the second term accounts for the difference $\log \Nh-\log \Nhpp$.
For example, the effect of doubling $J_\nu$ while keeping other parameters
fixed is to shift the LOX$(\log \Nh)$ curves diagonally
by the vector (0.13,0.13), indicated by the solid arrow in Figure~3.
The magnitude of this shift is small, and the direction is nearly parallel
to the LOX curve itself.  The same shift would apply to Figure~4.

If we keep $J_\nu$ fixed but change $\Omega_b$ by a factor $q$, then
absorber densities $n_{\rm H}$ change by $q$ and column densities by $q^2$
(because the hydrogen neutral fraction is itself proportional to $n_{\rm H}$).
An absorber that previously had column density $\Nhp$ now has
$\Nh = q^2\Nhp$.  The ionization parameters are the same as those in
an absorber that previously had column density $\Nhpp=q^{1.43}\Nhp=q^{-0.57}\Nh.$
With the same reasoning as before, we find
\begin{equation}
{\rm LOX}(q\Omega_b,\log \Nh) = {\rm LOX}(\Omega_b,\log \Nh+0.57\log q) -0.57\log q,
\qquad {\rm fixed}~J_\nu.
\label{scal2}
\end{equation}

Since the observational constraints on $D_A$ are better than those on 
the ionizing background intensity, a factor $q$ change in $\Omega_b$
should really be accompanied by a factor $q^2$ change in $J_\nu$,
so that $D_A$ and the HI column densities of absorbers remain the same.
The net change is to increase $\Gamma/n_{\rm H}$ by a factor of $q$ at fixed $\Nh$.
An absorber with column density $\Nh$ has the same ionization parameters
as an absorber that previously had column density $\Nhpp = q^{-1.43}\Nh$.
Thus,
\begin{equation}
{\rm LOX}(q\Omega_b,q^2 J_\nu,\log \Nh) = {\rm LOX}(\Omega_b,J_\nu,\log \Nh-1.43\log q) +1.43\log q,
\qquad {\rm fixed}~D_A.
\label{scal3}
\end{equation}
For example, the dashed arrow in Figure~3 shows the effect of increasing 
$\Omega_b$ from the value $\Omega_b=0.03$ used in our simulation to
the value $\Omega_b=0.047$ favored (for $h=0.65$) by the low [D/H] value
of Tytler et al.\ (1996).

Plausible variations in the assumed gas reionization history can alter
the temperature of the low density IGM by factors $\sim 2-3$
(Miralda-Escud\'e \& Rees 1994; Hui \& Gnedin 1997).  Because the 
HI recombination coefficients are $\propto T^{-0.7}$ for 
$T \sim 10^4\;$K, a factor $q$ change in gas temperatures requires
a factor $q^{-0.7}$ change in $J_\nu$ to keep $D_A$ fixed.
From the scaling relation~(\ref{scal1}), we see that the impact
of such a temperature change would be small.

To estimate the evolution of the LOX with redshift, we can appeal to the
argument of Hernquist et al.\ (1996) and Miralda-Escud\'e et al. (1996)
that the primary driver of evolution in the \lya forest at high
redshift is the expansion of the universe, which changes the cosmic
mean density in proportion to $(1+z)^3$.  Non-linear gravitational 
evolution changes the overdensity field $\rho/{\overline\rho}$, but this
evolution has a smaller impact on the forest than the overall expansion,
at least for low column density absorbers.  Matching the observed
evolution of $D_A$ requires an HI photoionization rate $\Gamma_{\rm HI}$
that is roughly constant between $z=4$ and $z=2$ (Hernquist et al.\ 1996;
Miralda-Escud\'e et al.\ 1996; Rauch et al.\ 1997), so if we ignore
changes to the shape of the spectrum we can take $J_\nu$ to be
constant over this interval.  A factor of two {\it decrease} in 
$(1+z)^3$ --- roughly the change between $z=4$ and $z=3$ or
between $z=3$ and $z=2$ --- has the same effect as a factor of two
{\it increase} in $J_\nu$ at a given redshift, illustrated by
the solid arrow in Figure~3.  While this argument is only approximate,
it is enough to show that the LOX should not vary strongly with
redshift over the range $z=2-4$, except in the case of ionic species that
are affected by the change in the spectral shape of $J_\nu$.

Structure formation models that have a power spectrum amplitude similar
to that of $\Lambda$CDM on Mpc scales at $z=3$ will predict similar spatial
and density structure in the diffuse IGM.  (Models with very different
power spectra will of course predict different structure, but
such models may soon be ruled out by the \lya forest data themselves
[see, e.g., discussions by Rauch et al.\ 1997; Gnedin 1997b; 
Croft et al.\ 1998].)  If the baryon density $\Omega_b h^2$ is held
fixed, then the most important effect of changing the cosmological
model is to alter the expansion rate $H(z)$.  Matching $D_A$
requires $\Gamma_{\rm HI} \propto H^{-1}(z)$ (see, e.g., 
equation~[2] of Rauch et al.\ 1997), so halving $H(z)$ has the
same impact as doubling $J_\nu$.  In practice the difference
between $H(z)$ values in currently popular models at $z\sim 3$
is less than a factor of two, so the expected impact is small.

We conclude that our quantitative
predictions for the LOX are likely to hold rather generally in
the cosmological picture of LYFAs, at least for column densities
$\Nh \la 10^{15}\; {\rm cm}^{-2}$, where the scaling arguments
used in this Section are most secure.
Our comparison of two numerical simulation models, a standard CDM
model and a $\Lambda$CDM model, supports this claim, and it does
not show any major changes to the LOX predictions
even for higher column densities.
Of course, the LOX is always directly affected by the metallicity
of the absorbing gas (equation~[\ref{lox}]), and
the insensitivity to cosmology implies that absorption
observations can therefore yield robust estimates of the metallicity
in different regions of the IGM.  Ionization corrections are
essential in the interpretation of such measurements, but
the corrections are not strongly dependent on the assumed cosmological model.

\section{Discussion}
In this paper we have introduced the LOX, a measure 
that allows a comprehensive selection of the metal lines that are
useful for testing the detailed predictions of models for LYFAs. We have 
applied this method to a particular model, a cosmological $\Lambda$CDM simulation,
and we have quantitatively discussed the dependence of the results on the model parameters
$J_{\nu}$, $\Omega _b$, and $z$. We have also performed this analysis for
a different cosmological simulation, a standard CDM model, but the results differ
negligibly from those of the $\Lambda$CDM model.

We find that very few detectable metal lines are associated with the low to moderate density regions
of the IGM that give rise to absorbers with $\log{N_{\rm HI}} \simlt 15$. Most of the
baryons in the Universe at $z=3$ are believed to reside in this part of the IGM
(e.g., Zhang et al. 1997), and the OVI(1032,1038) doublet is found to be the
best probe of the metallicity in these regions. Other strong O lines, such as OV(630),
OIV(788), and OIII(833) are technically difficult to observe, and have yet to be detected. 
The CIII(977) and SiIII(1206) lines are the
strongest metal lines in high column density LYFAs.
The CIV(1548) line is the strongest metal line with $\lambda _r > 1216 {\rm \AA}$ for all LYFAs
with $\log{N_{\rm HI}} \simlt 17$.
While the specific values of the LOX as a function of column density will depend
on redshift and on the adopted cosmological model and radiation field,
we have argued that the qualitative conclusions presented above are insensitive to the choice
of cosmological model of high redshift LYFAs, and they should hold generally within this scenario.

The relatively uncommon absorbers with $\log{N_{\rm HI}} \simgt 16$ should offer
the strongest test of cosmological simulations. These systems should have more than a 
dozen observable metal lines associated with them (for ${\rm [O/H]} \sim -2.5$).
The relative strengths of these lines will depend on the assumed radiation field
and abundance pattern, so a first check would be whether it is possible to make the
line strengths in the model spectra match those of lines associated with high
column density LYFAs for reasonable values of model parameters. Then one should
make a detailed examination of the relative locations in velocity space between
absorption components of individual species. 
A detailed comparison to observations will ultimately determine if
the resolution of the simulations is adequate in these high-density regions, 
and to what extent local dynamical effects associated with the production of metals
can be neglected. There are some hints of discrepancies between models and observations
for these high column density systems.
For example, the models in some 
cases predict the velocity components of CIV and CII absorption lines to differ,
whereas there seems to be some new observational evidence that CII and CIV components
are found at the same positions (A. Boksenberg, private communication).

Cosmological simulations depict the typical low column density LYFAs
as simple structures: low density, smooth, and governed by the
straightforward physics of gravity, cosmic expansion, and photoionization.  Studies of
absorption along parallel lines of sight provide direct observational
support for this point of view, especially recent HIRES observations
of gravitationally lensed QSOs that set stringent lower limits to
the scale of any substructure within low column density absorbers
(Rauch 1997; see also Smette et al. 1992, 1995).
The simulations do not predict a sudden change in LYFA properties
at any particular column density, but they do predict a steady
increase of gas density with $N_{\rm HI}$ (Fig.\ 1).
For ${\rm log}N_{\rm HI} \simgt 16$, the typical absorbers are
partially collapsed structures, and gas within them has begun to 
experience radiative cooling.  In this regime, the finite mass
and spatial resolution of the simulations may limit the accuracy
of the results, and local astrophysical processes such as star
formation and supernova explosions might influence the structure
of these systems.  The standard CDM model studied here produces
only about 1/3 of the observed number of Lyman limit systems
(${\rm log}N_{\rm HI} \geq 17.2$), even though it matches
the abundance of much stronger, damped \lya absorbers
quite well (Katz et al.\ 1996b; Gardner et al.\ 1997).
While further work is needed to assess the sensitivity of this
result to cosmological parameters --- especially the baryon
density parameter $\Omega_b$ --- the discrepancy of numbers
suggests that a population of absorbers unresolved by the 
simulations may become important at column densities
${\rm log}N_{\rm HI} \sim 17$.

The high column density LYFAs in the cosmological simulations
represent the transition between the weak IGM fluctuations
traced by low column density lines and the cold gas
concentrations in virialized dark halos traced by damped \lya
absorption.  The high-LOX lines listed in Table~1 can reveal many
details of the structure and physical conditions in these absorbers,
testing the accuracy of the simulations in this regime, providing clues 
to the nature of any additional absorber populations, and capturing 
snapshots of gas as it makes its way from the IGM into high
redshift galaxies.

\section*{Acknowledgements}
We thank the referee for drawing our attention to the importance
of the absorption lines in the extreme UV.
UH acknowledges support by a postdoctoral research 
grant from the Danish Natural Science Research Council. 
This work was supported by the NSF under grant
ASC93-18185 and the Presidential Faculty Fellows Program
and by NASA under grants NAG5-3111, NAG5-3525, and NAG5-3922.
Computing support was provided by the San Diego Supercomputer Center.

\clearpage

\begin{table}
\caption{LOX sorted lines for five different values of $\log{N_{\rm HI}}$}
\begin{tabular}{|l|r|l|r|l|r|} \hline \hline
\multicolumn{2}{|c|}{{ $\log{N_{\rm HI}}={\bf 13}$}} & 
\multicolumn{2}{c|}{\bf 16} &
\multicolumn{2}{|c|}{\bf 17}  \\ \hline
Line & LOX & Line & LOX & Line & LOX \\ \hline
OVI(1032) & 0.21 &  CIII(977) & 2.52 & CIII(977)  & 2.98 \\
OVI(1038) & -0.08 & CIV(1548) & 2.32 & OIII(833) & 2.43 \\
OV(630) & -0.55 & OIV(788) & 2.07 & OIII(702) & 2.39\\ 
NV(1238) & -0.90 & OIV(554) & 2.07 & SiIII(1207) & 2.35 \\
CIV(1548) & -0.94 & CIV(1551) & 2.02 & OIII(507) & 2.24 \\ \cline{1-2}
\multicolumn{2}{|c|}{\bf 14} & OIII(833) & 1.95 & OIII(306) & 2.21 \\ \cline{1-2}
OVI(1032) & 1.03 & NIV(765) & 1.92 & CIV(1548) & 2.16 \\
OV(630) & 0.74 & OIII(702) & 1.91 & CIV(1551) & 1.86 \\
OVI(1038) & 0.74 & OV(630) & 1.82 & OIII(303) & 1.74 \\
CIV(1548) & 0.33  & OIV(553) & 1.77 & NIV(765) & 1.69 \\
NV(1239) & 0.13  & OIII(507) & 1.76 & NIII(686) & 1.68 \\
CIV(1551) & 0.03  & OIII(306) & 1.73 & NIII(990) & 1.66 \\
NV(1243)  & -0.17 & OIV(238) & 1.69 & CIII(386) & 1.66 \\
OIV(788)  & -0.29  & NIII(990) & 1.29 & OIII(374) & 1.62 \\ \cline{1-2}
\multicolumn{2}{|c|}{\bf 15} & NV(1239) & 0.81 & SiIV(1394) & 1.54 \\ \cline{1-2}
OV(630) & 1.81 & SiIII(1207) & 0.80 & SIII(678) & 1.52 \\
CIV(1548) & 1.42 & NV(1243) & 0.52 & OIII(267) & 1.49 \\
OVI(1032) & 1.39 & OVI(1032) & 0.49 & SiIV(1403) & 1.25 \\
OIV(788) & 1.26 & CII(1335) & 0.47 & MgII(2796) & 1.16 \\
OIV(554) & 1.26 & SiIV(1394) & 0.40 & AlII(1671) & 1.10 \\
CIV(1551) & 1.12 & SIV(1063) & 0.30 & CII(1335) & 1.08 \\
NIV(765) & 1.11 & CII(1036) & 0.23 & SiII(1260) & 1.03 \\
OVI(1038) & 1.09 & OVI(1038) & 0.19 & MgII(2804) & 0.85 \\
OIV(553) & 0.96 & SiIV(1403) & 0.10 & CII(1036) & 0.84 \\
NV(1239) & 0.90 & SIII(1013) & 0.01 & SiII(1193) & 0.68 \\
OIV(238) & 0.88 &  &  & SiII(1190) & 0.38 \\
OIV(608) & 0.82 &  &  & AlIII(1855) & 0.35 \\
CIII(977) & 0.77 &  & & SIV(1063) & 0.32 \\
NV(1243)  & 0.60 &  &  & SIII(1013) & 0.27 \\
NeV(568) & 0.27 & & & SiII(1527) & 0.25 \\
OIII(833) & 0.26 & & & SIII(1190) & 0.19 \\
OIII(702) & 0.22 & & & AlIII(1863) & 0.05 \\
\hline \hline
\end{tabular}
\end{table}

\figcaption[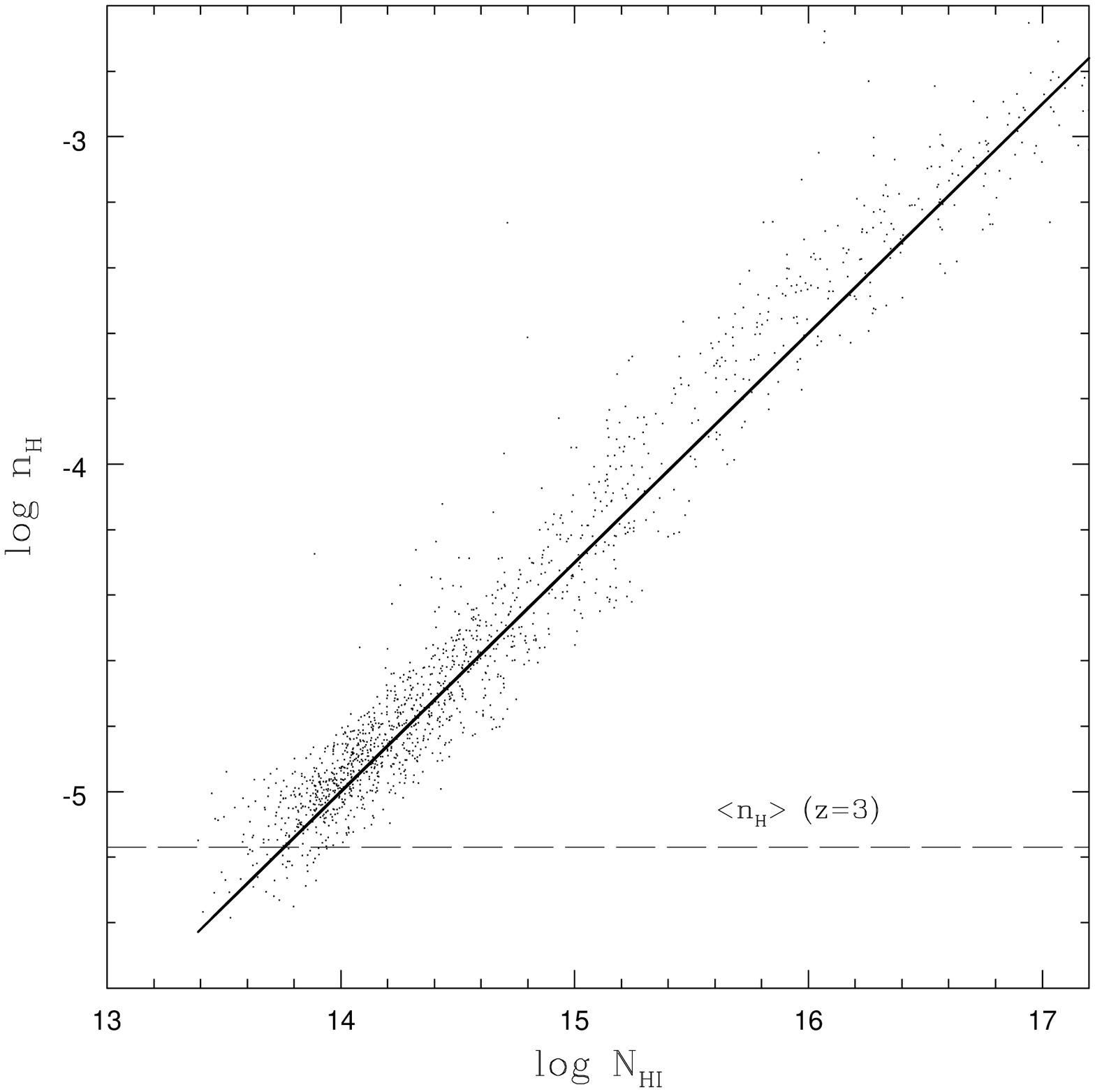]{Maximum total hydrogen number density in units of  $\log{({\rm cm}^{-3})}$ 
versus the HI column density in units of $\log{({\rm cm}^{-2})}$ for 1456 HI absorption features 
along 480 lines of sight at $z=3$ in our cosmological model. The solid line shows the fit 
$\log{n_{\rm H}}=-14.8+0.7\log{N_{\rm HI}}$. 
Also shown is the mean hydrogen density at $z=3$ (dashed line). 
\label{figcolnh}
}

\figcaption[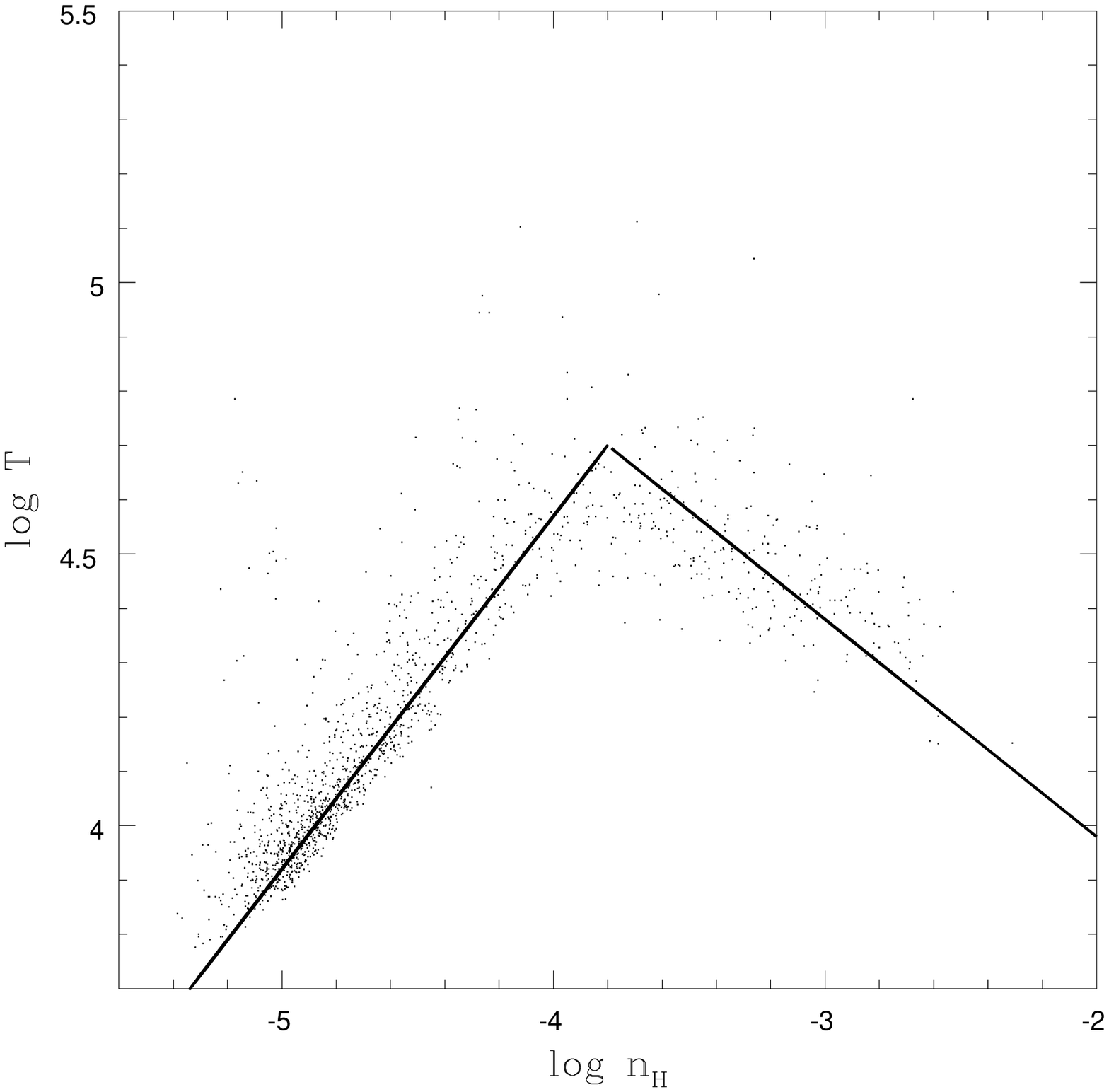]{Temperature vs. hydrogen number density for the same density peaks 
as plotted in figure 1. The shown fit is described by $T=7.17+0.65\log{n_{\rm H}}$ for
$\log{T}<-3.8$ and $T=3.18-0.4\log{n_{\rm H}}$ for $\log{T}>-3.8$.
\label{figtnh}
}

\figcaption[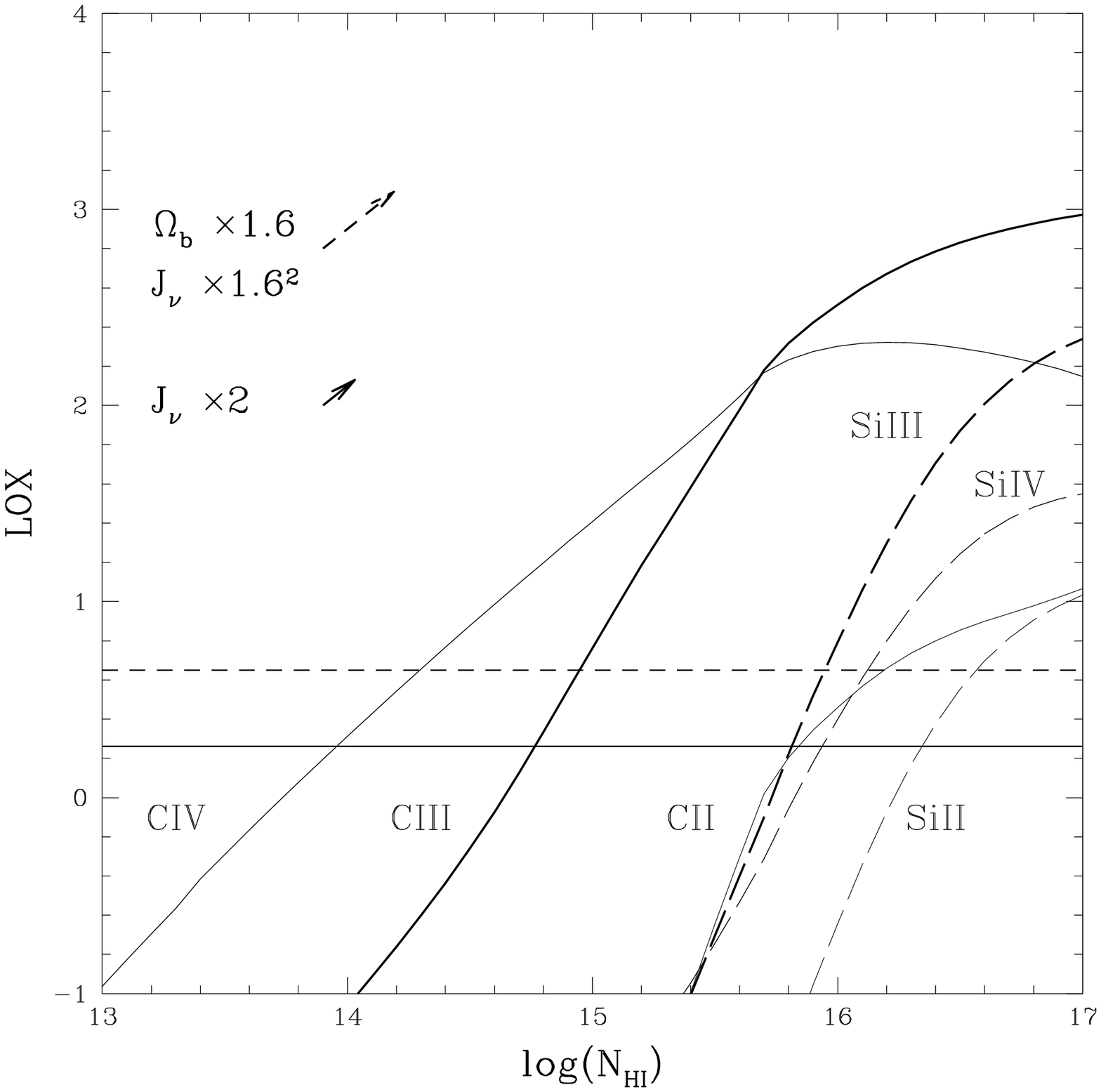]{LOX($N_{\rm HI}$) for CIV(1548), CIII(977), and CII(1335) (solid
curves) and SiIV(1394), SiIII(1207), and SiII(1260) (dashed curves) 
at $z=3$, assuming ${\rm [O/H]} = -2.5$. 
The horizontal lines are approximate detection limits
within the \lya forest (dashed line) and redwards of the forest (solid line) in 
Songaila and Cowie's spectrum of Q1422+231 (equation (6)). Arrows indicate the effects of 
doubling the radiation intensity (solid arrow) and of changing $\Omega _b h^2$ from 0.0125 to
0.02, as discussed in Section 4.
}

\figcaption[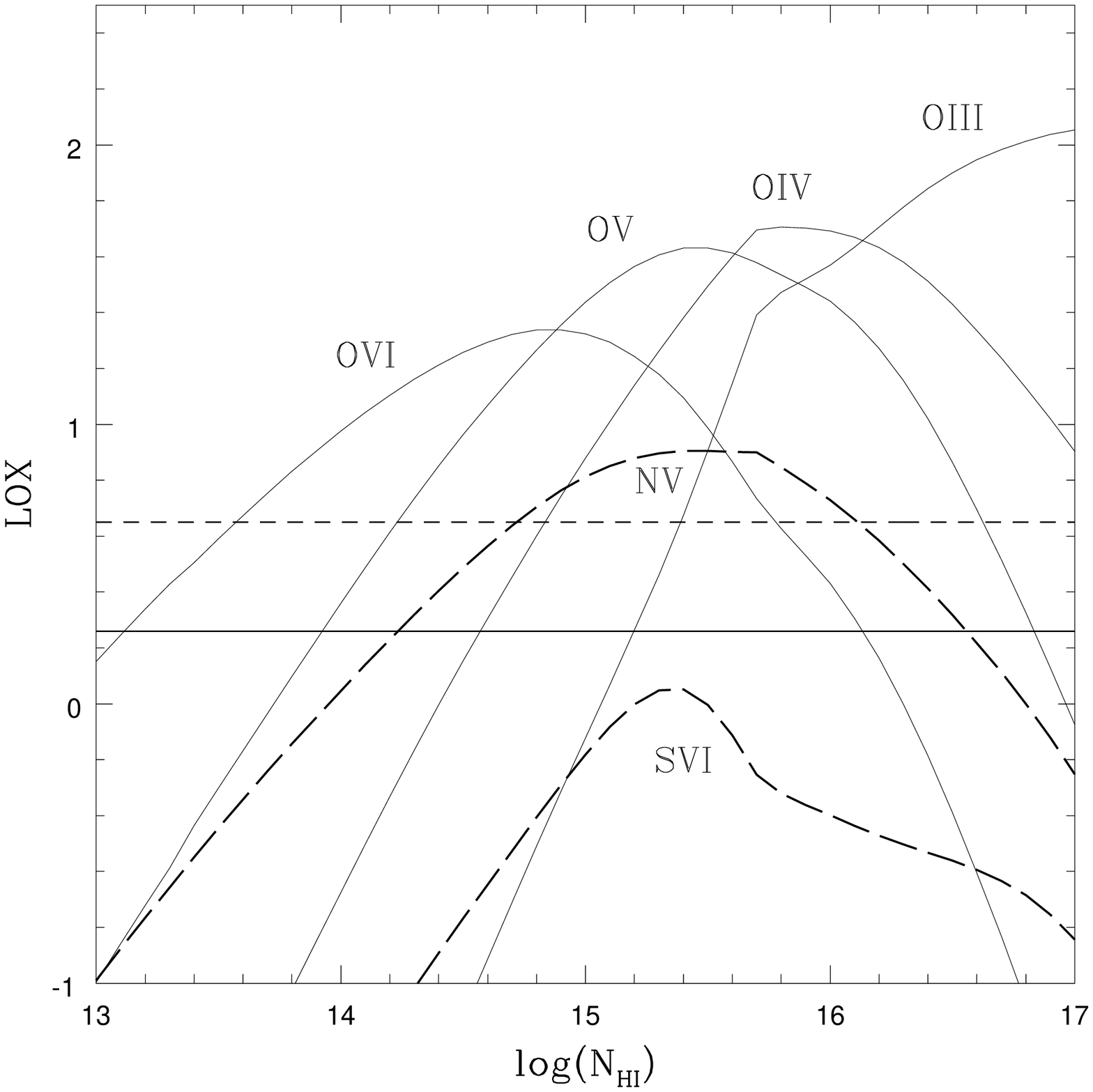]{LOX($N_{\rm HI}$)for OVI(1032), OV(630), OIV(788), and
OIII(833) (solid curves), and NV(1243), and SVI(933) (dashed curves).
}

\clearpage

\plotone{colnh.ps}

\clearpage

\plotone{tnh.ps}

\clearpage

\plotone{lox_CSi.ps}

\clearpage

\plotone{lox_NOS.ps}

\vfill\eject


\clearpage

\begin{thebibliography}{}


\bibitem[{Bi \& Davidsen }1997]{bi97}
    Bi, H.G., \& Davidsen, A. 1997, \apj , 479, 523
\bibitem[Cen et al. 1994]{cen94} Cen, R., Miralda-Escud\'e, J.,
    Ostriker, J.P., \& Rauch M. 1994, \apj, 437, L9
\bibitem[]{} Churchill, C. 1997, Ph.D. Thesis, Univ. of California,  Santa Cruz.
\bibitem[]{} Cowie, L.L., Songaila, A., Kim, T.-S., \& Hu, E.M. 1995,
    AJ., 109, 1522 
\bibitem[]{} Croft, R.A.C., Weinberg, D.H., Katz, N., \& Hernquist, L. 1997,
   \apj, in press
\bibitem[]{} Croft, R.A.C., Weinberg, D.H., Katz, N., \& Hernquist, L. 1998,
   \apj, submitted
\bibitem[]{} Danks, A., Woodgate, B., Kimble, R., Bowers, C., Grady, J.,
             Kraemer, S., Kaiser, M.E., Meyer, W., Hood, D. \& van Houten, C. 1996,
             in Science with the HST-II, Space Telescope Science Institute,
             eds. Benvenuti, Macchetto \& Schreier.
\bibitem[]{} Dav\'{e}, R., Hernquist, L., Weinberg, D.H., \& Katz, N. 1997,
    \apj, 477, 21
\bibitem[]{} Ferland, G.J., 1996, University of Kentucky, Department of
    Astronomy, Internal report
\bibitem[]{} Gardner, J.P., Katz, N., Hernquist, L., \& Weinberg, D.H. 1997,
    \apj, in press (astro-ph/9609072)
\bibitem[Gnedin 1997a]{gnedin97}
Gnedin, N. Y. 1997a, \mnras, submitted (astro-ph/9707257)
\bibitem[Gnedin 1997b]{gnedin97}
Gnedin, N. Y. 1997b, \mnras, submitted (astro-ph/9706286)
\bibitem[]{} Gnedin, N. Y. \& Ostriker, J. P., \apj, submitted, astro-ph/9612127
\bibitem[]{} Green, J., Shull, J.M., Morse, J. et al. 1997,
	     proposal for "Cosmic Origins Spectrograph", selected by NASA for
             Hubble Space Telescope 2002 Refurbishment Mission.
\bibitem[]{} Grevesse, N. \& Anders, E. 1989, { Cosmic Abundances of Matter},
AIP Conf. Proceedings 183,1.
\bibitem[]{} Gunn, J.E. \& Peterson, B. A. 1965, \apj, 142, 1633
\bibitem[HM]{haa96} Haardt, F. \& Madau, P. 1996, \apj, 461, 20 
\bibitem[HM]{} Haardt, F. \& Madau, P. 1997, in prep. 
\bibitem[]{} Haehnelt, M.G., Steinmetz, M., \& Rauch, M. 1996, \apj, 465, L65  
\bibitem[Hernquist \& Katz 1989]{her89} Hernquist, L. \& 
    Katz, N. 1989, \apjs, 70, 419
\bibitem[]{} Hellsten, U., Dav\'{e}, R., Hernquist, L., Weinberg, D. \& Katz, N. 1997,
             \apj, vol. 487
\bibitem[HKWM]{her96} Hernquist, L., Katz, N., Weinberg, D.H., 
    \& Miralda-Escud\'e, J. 1996, \apjl, 457, L51 
\bibitem[]{} Hui, L., Gnedin, N., 1997, \mnras, submitted, astro-ph/9612232
\bibitem[]{} Katz, N., Weinberg, D.H., \& Hernquist, L. 1996a,
             ApJS, 105, 19
\bibitem[]{} Katz, N., Weinberg D.H., Hernquist, L., \& Miralda-Escud\'e, J. 1996b, 
             \apj, 457, L57
\bibitem[]{} Lu, L. 1991,
	     \apj, 379, 99
\bibitem[]{} Meyer, D.M. \& York, D.G. 1987,
	     \apj, 315, L5
\bibitem[]{} Miralda-Escud\'{e}, J., Cen, R., Ostriker, J.P. \& Rauch, M. 1996,
	     \apj, 471, 582
\bibitem[Miralda-Escud\'e \& Rees 1994]{miralda94}
Miralda-Escud\'e J., \& Rees, M. J. 1994, \mnras, 266, 343
\bibitem[]{} Petitjean, P., M\"{u}cket, J.P. \& Kates, R.E. 1995,
        A\&A, 295, L9   
\bibitem[]{} Pettini, M., Lipman, K., \& Hunstead, R.W. 1995,
    \apj, 451, 100 
\bibitem[PRS]{pre93} Press, W.H., Rybicki, G.B., \&
    Schneider, D.P. 1993, \apj, 414, 64 
\bibitem[Rauch 1997]{} Rauch, M. 1997, in Proc. of the 13th IAP
    Colloquium, Structure and Evolution of the IGM from QSO Absorption
    Line Systems, eds. P. Petitjean \& S. Charlot, (Paris: Nouvelles
    Fronti\`eres), astro-ph/9709129
\bibitem[]{} Rauch, M., Haehnelt, M.G., \& Steinmetz, M. 1996, \apj, 481, 601.
\bibitem[]{} Rauch, M., Miralda-Escud\'{e}, J., Sargent, W.L.W., Barlow, T.A.,
             Weinberg, D.H., Hernquist, L., Katz, N.S. \& Ostriker, J.P. 1997,
             \apj, in press, astro-ph/9612245. 
\bibitem[]{} Reimers, D., K\"ohler, S., Wisotzki, L., Groote, D.,
             Rodriguez-Pascual, P., \& Wamsteker, W. 1997,
	     A \& A, 327, 890.
\bibitem[]{} Seljak, U. \& Zaldarriaga, M., 1996,
	     \apj, 469, 437
\bibitem[]{} Smette, A., Surdej, J., Shaver, P. A., Foltz, C. B., 
	     Chaffee, F. H., Weymann, R. J., Williams, R. E., Magain, P. 1992, 
	     \apj, 389, 39
\bibitem[]{} Smette, A., Robertson, J. G., Shaver, P A., Reimers, D.,
             Wisotzki, L., Koehler, T. 1995, A\&AS, 113, 199
\bibitem[Smoot et al.\ 1992]{smoot92}
             Smoot, G. F., et al.\ 1992, \apjl, 396, L1
\bibitem[]{} Songaila, A. \& Cowie, L. L. 1996,
    AJ, 112, 335 [SC]
\bibitem[]{} Spitzer, L. Jr., 1978, {Physical Processes in the Interstellar
             Medium}, Wiley-Interscience.
\bibitem[]{} Tytler, D., Fan, X.-M., and Burles, S. 1996,
             Nature, 381, 207
\bibitem[]{} Verner, D.A., Barthel, P.D., \& Tytler, D. 1994,
	     A\&AS, 108, 287
\bibitem[]{} Wheeler, J.C., Sneden, C., \& Truran, J.W. 1989,
    ARA \& A, 27, 279
\bibitem[]{} Womble, D.S., Sargent, W.L.W., \& Lyons, R.S. 1995, 
     in {Cold Gas at High Redshift}, eds. M. Bremer et al., Kluwer 1996,
     (astro-ph/9511035).
\bibitem[Zhang, Anninos, \& Norman 1995]{zha95} Zhang, Y., Anninos, P.,
    \& Norman, M.L. 1995, \apjl, 453, L57
\bibitem[]{} Zhang, Y., Meiksin, A., Anninos, P., Norman, M.L., 1997,
ApJ, in press (astro-ph 9706087).
\bibitem[Zuo \& Lu 1993]{zuo93}
Zuo, L., \& Lu, L. 1993, \apj, 418, 601
\end{thebibliography}
\end{document}